\documentstyle[12pt,epsf]{article}
\begin{document}

\title{Interaction of a vortex ring with the free surface of ideal fluid}
\author{V.P. Ruban 
\footnote{Permanent address: L.D.Landau Institute for Theoretical Physics,
 2 Kosygin str., 117334 Moscow, Russia. ~~~E-mail: ruban@itp.ac.ru}\\
 {\it Optics and Fluid Dynamics Department,}\\
 {\it Ris\o ~National Laboratory, DK-4000 Roskilde Denmark}}

\maketitle
\begin{abstract}
The interaction of a small vortex ring with the free surface of a perfect fluid 
is considered. In the frame of the point ring approximation the asymptotic
expression for the Fourier-components of radiated surface waves is obtained
in the case when the vortex ring comes from infinity and has both horizontal 
and vertical components of the velocity. The non-conservative corrections 
to the equations of motion of the ring, due to Cherenkov radiation, are derived.
\end{abstract}

\section{Introduction}

The study of interaction  between vortex structures in a fluid and 
the free surface is important both from practical and theoretical points of 
view. In general, a detailed investigation of this problem is very hard. 
Even the theories of potential surface waves 
and the dynamics of vortices in an infinite space taken 
separately still have a lot of unsolved fundamental problems on their own.
Only the consideration of significantly simplified models can help us 
to understand the processes which take place in the combined system.

In many cases it is possible to neglect the compressibility of the fluid 
as well as the energy dissipation.
Therefore the model of ideal homogeneous incompressible fluid is very
useful for hydrodynamics. Because of the conservative nature of this model 
the application of the well developed apparatus of Hamiltonian dynamics
becomes possible \cite{Arnold} \cite{ZK97}.  
An example of effective use of the Hamiltonian
formalism in hydrodynamics is the introduction of canonical
variables for investigations of potential flows of perfect fluids with a 
free boundary. V.E.Zakharov showed
at the end of the sixties \cite{Z68} that the surface shape $z=\eta(x,y,t)$
and the value of the velocity potential $\psi(x,y,t)$ on the surface can be
considered as generalized coordinate and momentum, respectively.

It is important to note that a variational formulation
of Hamiltonian dynamics  in many cases allows to obtain good finite-dimensional
approximations which reflect the main features of the behavior of the original system.
There are several possibilities
for a parameterization of non-potential flows of perfect fluid by
some variables with dynamics determined by a variational principle.
All of them are based on the conservation of the topological characteristics
of vortex lines in ideal fluid flows which follows from the freezing-in of 
the vorticity field ${\bf \Omega }({\bf r},t)=\mbox{curl}\,{\bf v}({\bf r},t)$.
In particular, this is the representation of the vorticity 
by Clebsch canonical variables $\lambda$ and $\mu$ 
\cite{Lamb} \cite{ZK97}
$$
{\bf \Omega }({\bf r},t)=[\nabla\lambda\times\nabla\mu]
$$
However, the Clebsch representation can only describe flows with 
a trivial topology (see, e.g., \cite{KM80}). 
It cannot describe flows with linked vortex lines. Besides, the variables 
$\lambda$ and $\mu$ are not suitable for the study of localized vortex 
structures like vortex filaments. In such cases it is more convenient to use 
the parameterization of vorticity in terms of vortex lines and
consider the motion of these lines \cite{Berdichevsky},\cite{KR98}, 
even if the global
definition of canonically conjugated variables is impossible due to 
topological reasons.

This approach is used in the present article to describe the interaction
of deep (or small) vortex rings of almost ideal shape 
in the perfect fluid with the free surface. 
In the case under consideration the main interaction of the vortex
rings with the surface can be described as the  dipole-dipole interaction 
between "point" vortex rings and their "images". Moving rings
interact with the surface waves, leading to radiation due to the Cherenkov effect. 
Deep rings disturb the surface weakly, so the influence of the surface
can be taken into account as some small corrections in the equations of motion
for the parameters of the rings.

In Sec.2 we discuss briefly general properties of vortex line dynamics, 
which follow from the freezing-in of the vorticity field.
In Sec.3 possible simplifications of the model are made 
and the point ring approximation is introduced.
In Sec.4 the interaction of the ring with its image is considered.
In Sec.5 we calculate the Fourier-components of Cherenkov surface waves 
radiated by a moving vortex ring 
and determine the non-conservative corrections caused by the 
interaction with the surface for the vortex ring equations of motion.

\section{Vortex lines motion in perfect fluid}

It is a well known fact that the freezing-in of the vorticity lines 
follows from the Euler equation for ideal fluid motion
$$
{\bf\Omega}_t=\mbox{curl}\,[{\bf v}\times{\bf\Omega}], \qquad
{\bf v}=\mbox{curl}^{-1}\,{\bf\Omega }
$$
Vortex lines are transported by the flow \cite{Arnold},\cite{Lamb},\cite{LL6}. 
They do not appear or disappear, 
neither they intersect one another in the process of motion.
This property of perfect fluid flows is general for all Hamiltonian systems of 
the hydrodynamic type.
For simplicity, let us consider temporally the incompressible fluid without 
free surface in infinite space. The dynamics of the system is specified
by a basic Lagrangian $L[{\bf v}]$, which is a functional of the solenoidal velocity
field. The relations between the velocity ${\bf v}$,
the generalized vorticity ${\bf\Omega}$, the basic Lagrangian $L[{\bf v}]$ and
the Hamiltonian ${\cal H}[{\bf\Omega}]$ are the following \cite{R99}\footnote{
For the ordinary ideal hydrodynamics in infinite space the basic Lagrangian is
$$
L_{Euler}[{\bf v}]=\int \frac{{\bf v}^2}{2}d{\bf r}\qquad 
\Rightarrow \qquad {\bf\Omega}=\mbox{curl}\,{\bf v}
$$
The Hamiltonian in this case coincides 
with the kinetic energy of the fluid and in terms of the vorticity field it reads$$
{\cal H}_{Euler}[{\bf\Omega}]=
-1/2\int {\bf\Omega}\Delta^{-1}{\bf\Omega}\,d{\bf r}=
\frac{1}{8\pi}\int\!\!\int
\frac{{\bf\Omega}({\bf r}_1)\cdot{\bf\Omega}({\bf r}_2)}
{|{\bf r}_1-{\bf r}_2|}d{\bf r}_1d{\bf r}_2
$$
where $\Delta^{-1}$ is the inverse Laplace operator.

Another example is the basic Lagrangian of Electron Magneto-hydrodynamics
which takes into account the magnetic field created by the current of electron 
fluid through the motionless ion fluid.
$$
L_{EMHD}[{\bf v}]=\frac{1}{2}\int{\bf v}(1-\Delta^{-1}){\bf v}\,d{\bf r}\qquad 
\Rightarrow \qquad {\bf\Omega}=\mbox{curl}(1-\Delta^{-1}){\bf v}
$$
$$
{\cal H}_{EMHD}[{\bf\Omega}]=\frac{1}{2}\int 
{\bf\Omega}(1-\Delta)^{-1}{\bf\Omega}\,d{\bf r}=
\frac{1}{8\pi}\int\!\!\int
\frac{e^{-|{\bf r}_1-{\bf r}_2|}}{|{\bf r}_1-{\bf r}_2|}
{\bf\Omega}({\bf r}_1)\cdot{\bf\Omega}({\bf r}_2)d{\bf r}_1d{\bf r}_2
$$

The second example shows that the relation between the velocity and the
vorticity can be more complex than in usual hydrodynamics.
}
\begin{equation}
{\bf\Omega}=\mbox{curl}\left(\frac{\delta L}{\delta{\bf v}}\right)\qquad
\Rightarrow \qquad {\bf v}={\bf v}[{\bf\Omega}]
\end{equation}
\begin{equation}
{\cal H}[{\bf\Omega}]=
\left(\int{\bf v}\cdot\left(\frac{\delta{ L}}{\delta{\bf v}}\right)d^3{\bf r}
-{L}[{\bf v}]\right)\Big|_{{\bf v}={\bf v}[{\bf\Omega}]}
\end{equation}
\begin{equation}
{\bf v}=\mbox{curl}\left(\frac{\delta{\cal H}}{\delta{\bf\Omega}}\right)
\end{equation}
and the equation of motion for the generalized vorticity is
\begin{equation}\label{Ham}
{\bf\Omega}_t=\mbox{curl}\,
[\mbox{curl}\,(\delta{\cal H}/\delta{\bf\Omega})\times{\bf\Omega}]
\end{equation}
This equation corresponds to the transport of frozen-in vortex lines by the 
velocity field. In this process all topological invariants \cite{MonSas} 
of the vorticity field are conserved. 
The conservation of the topology can be expressed by 
the following relation \cite{KR98}
\begin{equation}\label{OmegaR}
{\bf \Omega }({\bf r},t)=\int \delta({\bf r}-{\bf R}({\bf a},t))
({\bf \Omega}_{0}({\bf a})\nabla _{{\bf a}}){\bf R}({\bf a},t)d{\bf a}=
\frac{({\bf \Omega}_{0}({\bf a})\nabla _{{\bf a}}){\bf R}({\bf a},t)}
{\mbox{det}\|\partial{\bf R}/\partial{\bf a}\|}
\Big|_{{\bf a}={\bf a}({\bf r},t)}
\end{equation}
where the mapping ${\bf R}({\bf a},t)$ describes the deformation of lines of
some initial solenoidal field ${\bf \Omega}_{0}({\bf r})$. Here
${\bf a}({\bf r},t)$ is the inverse mapping with respect to
${\bf R}({\bf a},t)$. 
The direction of the vector ${\bf b}$
\begin{equation}\label{b}
{\bf b}({\bf a},t)=
({\bf \Omega}_{0}({\bf a})\nabla _{{\bf a}}){\bf R}({\bf a},t)
\end{equation}
coincides with the direction of the vorticity field at the point 
${\bf R}({\bf a},t)$. The equation of motion for the mapping 
${\bf R}({\bf a},t)$ can be obtained with the help of the relation
\begin{equation}\label{OmegaRt}
{\bf\Omega}_t({\bf r},t)=
\mbox{curl}_{\bf r}\int
\delta({\bf r}-{\bf R}({\bf a},t))
[{\bf R}_t({\bf a},t)\times{\bf b}({\bf a},t)]
d{\bf a},
\end{equation}
which immediately follows from Eq.(\ref{OmegaR}). 
The substitution of Eq.(\ref{OmegaRt})
into the equation of motion (\ref{Ham}) gives \cite{KR_PRE99}
$$
\mbox{curl}_{\bf r}\left(\frac{{\bf b}({\bf a},t)\times
[{\bf R}_t({\bf a},t)-{\bf v}({\bf R},t)]}
{\mbox{det}\|\partial{\bf R}/\partial{\bf a}\|}\right)=0
$$
One can solve this equation by eliminating the $\mbox{curl}_{\bf r}$ operator.
Using the general relationship between variational derivatives 
of some functional $F[{\bf \Omega }]$
\begin{equation}\label{peresch}
\left[ {\bf b}\times \mbox{curl}\left( \frac{\delta F}{\delta {\bf \Omega }
({\bf R})}\right) \right] =\frac{\delta F}{\delta {\bf R}({\bf a})}
\Big|_{{\bf \Omega }_{0}}  
\end{equation}
it is possible to represent the equation of motion for ${\bf R}({\bf a},t)$ 
as follows
\begin{equation}
\left[ ({\bf \Omega }_{0}({\bf a})\nabla _{{\bf a}}){\bf R}({\bf a})\times
{\bf R}_{t}({\bf a})\right] =
\frac{\delta {\cal H}[{\bf \Omega }[{\bf R}]]}
{\delta {\bf R}({\bf a})}\Big|_{{\bf \Omega }_{0}}.  \label{main}
\end{equation}
It is not difficult to check now that the dynamics of the vorticity field with 
topological properties defined by ${\bf \Omega}_{0}$ in the infinite 
space is equivalent to the requirement of an extremum of the action 
($\delta S=\delta\int {\cal L}_{{\bf \Omega}_{0}}dt=0$) 
where the Lagrangian is \cite{KR98} 
\begin{equation}
{\cal L}_{{\bf \Omega}_{0}}=
\frac{1}{3}\int\Big(\left[ {\bf R}_{t}({\bf a})\times
{\bf R}({\bf a})\right]\cdot({\bf \Omega }_{0}({\bf a})\nabla _{{\bf a}})
{\bf R}({\bf a})\Big)d{\bf a} 
 - {\cal H}[{\bf \Omega }[{\bf R}]].  \label{LAGRANGIAN}
\end{equation}

In the simplest case, when all vortex lines are closed 
it is possible to choose new 
curvilinear coordinates $\nu_1,\nu_2,\xi$ in ${\bf a}$-space such that  
Eq.(\ref{OmegaR}) can be written in a simple form
\begin{equation}\label{lines}
{\bf \Omega }({\bf r},t)=\int_{\cal N}d^2\nu \oint \delta ({\bf r}-
{\bf R}(\nu,\xi,t)){\bf R}_{\xi}d\xi.  
\end{equation}
Here $\nu$ is the label of a line lying on a fixed two-dimensional manifold
${\cal N}$, and $\xi$ is some parameter along the line.
It is clear that there is a gauge freedom in the definition of $\nu$ and $\xi$.
This freedom is connected with the possibility of changing the longitudinal parameter 
$\xi=\xi(\tilde\xi,\nu,t)$ and also with the relabeling of $\nu$
\begin{equation}\label{nu_relabl}
\nu=\nu(\tilde\nu,t),\qquad
\frac{\partial(\nu_1,\nu_2)}
{\partial(\tilde\nu_1,\tilde\nu_2)}=1.
\end{equation}

Now we again consider the ordinary perfect fluid with a free surface.
To describe the flow entirely it is sufficient to
specify the vorticity field ${\bf \Omega }({\bf r},t)$ and the motion of
the free surface. Thus, we can use the shape 
${\bf R}(\nu,\xi,t)$ of the vortex lines as a new dynamic object instead of 
${\bf \Omega }({\bf r},t)$. It is important to note 
that in the presence of the free surface the equations of motion 
for ${\bf R}(\nu,\xi,t)$ follow from a variational principle
as in the case of infinite space. It has been shown 
\cite{KR99} that the Lagrangian for a perfect fluid, with vortices in its bulk 
and with a free surface, can be written in the form
\begin{equation}\label{Lagr1}
{\cal L}=\frac{1}{3} \int_{\cal N} d^2\nu 
\oint([{\bf R}_t\times{\bf R}]\cdot{\bf R}_{\xi})d\xi+
\int \Psi \eta_t d{\bf r}_{\bot} -{\cal H}[{\bf R},\Psi,\eta].
\end{equation}
The functions $\Psi({\bf r}_{\bot},t)$ and $\eta({\bf r}_{\bot},t)$
are the surface degrees of freedom for the system. $\Psi$ is the boundary
value of total velocity potential, which includes the part from vortices 
inside the fluid, 
and $\eta$ is the deviation of the surface from the horizontal plane.
This formulation supposes that vortex lines do not intersect the 
surface anywhere. In the present paper only this case is considered.

The Hamiltonian ${\cal H}$ in Eq.(\ref{Lagr1}) 
is nothing else than the total energy of the 
system expressed in terms of $[{\bf R},\Psi,\eta]$.

Variation with respect to 
${\bf R}(\nu,\xi,t)$ of the action defined by the Lagrangian (\ref{Lagr1}) 
gives the equation of motion for vortex lines in the form
\begin{equation}\label{transv}
[{\bf R}_\xi\times{\bf R}_t]=\frac{\delta{\cal H}[{\bf\Omega[R]},\Psi,\eta]}
{\delta{\bf R}} \; .
\end{equation}
This equation determines only the transversal component of ${\bf R}_t$
which coincides with the transversal component of the actual solenoidal 
velocity field. The possibility of solving Eq.(\ref{transv}) 
with respect to the time derivative ${\bf R}_t$ is closely connected with the 
special gauge invariant nature of the ${\cal H}[{\bf R}]$ dependence which results in 
$$
\frac{\delta{\cal H}}
{\delta{\bf R}}\cdot {\bf R}_\xi \equiv 0 \; .
$$ 

The tangential component of ${\bf R}_t$ with respect to vorticity direction 
can be taken arbitrary. 
This property is in accordance with the longitudinal gauge 
freedom. The vorticity dynamics does not depend on the choice of the 
tangential component. 

Generally speaking, only the local introduction of canonical variables for curve 
dynamics is possible. For instance, a piece of the curve 
can be parameterized by one of the three of Cartesian coordinates
$$
{\bf R}=(X(z,t),Y(z,t),z)
$$
In this case the functions $X(z,t)$ and $Y(z,t)$ are canonically conjugated
variables. Another example is the parameterization in cylindrical coordinates,
where variables $Z(\theta,t)$ and $(1/2)R^2(\theta,t)$ are canonically conjugated.

Curves with complicated topological properties need a general
gauge free description by means of a parameter $\xi$.

It should be mentioned for clarity that the conservation of all 
vortex tube volumes, reflecting the incompressibility of the fluid, 
is not the constraint in this formalism. 
It is a consequence of the symmetry of the Lagrangian (\ref{Lagr1}) with
respect to the relabeling (\ref{nu_relabl})  $\nu\to\tilde\nu$ \cite{R99}. 
Volume conservation follows from that
symmetry in accordance with Noether's theorem.
To prove this statement, we should consider such subset of relabelings
which forms a one-parameter group of transformations of the dynamical variables. 
For small values of the group parameter, $\tau$, the transformations are 
determined by a function of two variables $T(\nu_1,\nu_2)$ 
(with zero value on the boundary $\partial {\cal N}$) so that
\begin{equation}\label{transf}
{\bf R}(\nu_1,\nu_2,\xi)\to{\bf R}^\tau_T(\nu_1,\nu_2,\xi)=
{\bf R}\Bigg(\nu_1-\tau\frac{\partial T}{\partial \nu_2}+O(\tau^2),\,\,\,
\nu_2+\tau\frac{\partial T}{\partial \nu_1}+O(\tau^2),\,\,\,\xi\Bigg)
\end{equation}
Due to Noether's theorem, the following quantity is an integral of motion
\cite{DNF}
$$
I_T=\int_{\cal N} d^2\nu\oint
\frac{\delta{\cal L}}{\delta {\bf R}_t}\cdot
\frac{\partial{\bf R}^\tau_T}{\partial\tau}\Bigg|_{\tau=0} \!\!d\xi
=\frac{1}{3} \int_{\cal N} d^2\nu 
\oint[{\bf R}\times{\bf R}_\xi]\cdot({\bf R}_2 T_1-{\bf R}_1 T_2)d\xi
$$
After simple integrations in parts the last expression takes the form
\begin{equation}
I_T= \int_{\cal N} d^2\nu \oint T(\nu_1,\nu_2)
([{\bf R}_1\times{\bf R}_2]\cdot{\bf R}_\xi)d\xi=
\int_{\cal N}  T(\nu_1,\nu_2){\cal V}(\nu_1,\nu_2,t)d^2\nu
\end{equation}
where ${\cal V}(\nu_1,\nu_2,t)d^2\nu$ is the volume of an infinitely thin vortex 
tube with cross-section $d^2\nu$. It is obvious that actually the function
${\cal V}$ doesn't depend on time $t$ because the function $T(\nu_1,\nu_2)$ 
is arbitrary
\footnote{
If vortex lines are not closed but form a family of enclosed tori 
then the relabeling freedom is less rich. 
In that case one can obtain by the similar way the conservation laws 
for volumes inside closed vortex surfaces. Noether's theorem gives 
integrals of motion which depend on an arbitrary function of one variable 
$S(\zeta)$, where $\zeta$ is the label of the tori.
}.

\section{Point ring approximation}

In general case an analysis of the dynamics defined by the Lagrangian 
(\ref{Lagr1}) is too much complicated. We do not even have the exact expression for
the Hamiltonian ${\cal H}[{\bf R},\Psi,\eta]$ because it needs the 
explicit knowledge of the solution of the Laplace equation with a boundary value 
assigned on a non-flat surface. Another reason is the very high nonlinearity
of the problem.

In this paper we consider some limits where it 
is possible to simplify the system significantly. Namely, we will 
suppose that the vorticity is concentrated in several very thin vortex 
rings of almost ideal shape. For a solitary ring the perfect shape 
is stable for a wide range of vorticity distributions through the cross-section. 
This shape provides an extremum of the energy for given values of the volumes
of vortex tubes  and for a fixed momentum of the ring. As already 
mentioned, volume conservation follows from  Noether's theorem.
Therefore some of these quantities (those of which 
are produced by the subset of commuting transformations) 
can be considered as canonical momenta.
Corresponding cyclical coordinates describe the relabeling (\ref{nu_relabl})
of the line markers, which doesn't change the vorticity field. Actually these 
degrees of freedom take into account a rotation around the central line of the 
tube. This line represents the mean shape of the ring and we are interested in how
it behaves in time. For our analysis we don't need the explicit values 
of cyclical coordinates, but only the conserved volumes as parameters 
in the Lagrangian. 

A possible situation is  
when a typical time of the interaction with the surface and with other rings is
much larger then the largest period of oscillations corresponding to deviations 
of the ring shape from perfect one. Under this condition, excitations of all 
(non-cyclical) internal degrees of freedom are small during all the time, 
and a variational anzats completely disregarding them 
reflects the behavior of the system adequately. The circulations  
$$
\Gamma_n=\int_{{\cal N}_n}d^2\nu
$$ 
of the velocity for each ring don't depend on time. 
A perfect ring is described by the coordinate ${\bf R}_n$
of the center and by the vector ${\bf P}_n=\Gamma_n{\bf S}_n$, where
${\bf S}_n$ is an oriented area of the ring. We use in this work 
the Cartesian system of coordinates $(x,y,z)$, so that 
the vertical coordinate is $z$, and the unperturbed surface is at $z=0$.
The corresponding components of the vectors ${\bf R}_n$ and ${\bf P}_n$ are
$$
{\bf R}_n=(X_n,Y_n,Z_n),\qquad {\bf P}_n=(P_{xn},P_{yn},P_{zn})
$$
It is easy to verify that the 
vectors ${\bf P}_n$ are canonically conjugated momenta for the coordinates 
${\bf R}_n$. To verify that we can parameterize the shape of each
vortex line in the following manner
\begin{equation}
{\bf R}(\xi,t)=\sum_{m=-M}^{M}{\bf r}_m(t)e^{im\xi},
\qquad {\bf r}_{-m}=\bar{\bf r}_m
\end{equation}
Here ${\bf r}_m(t)$ are complex vectors. Substituting this into the first term
of the Lagrangian (\ref{Lagr1}) gives
$$
\frac{1}{3}\oint([{\bf R}_t\times{\bf R}]{\bf R}_{\xi})d\xi=
2\pi i\dot{\bf r}_0([{\bf r}_{-1}\times{\bf r}_1]+
2[{\bf r}_{-2}\times{\bf r}_2]+\dots)+
$$
\begin{equation}
+\frac{d \{...\}}{dt}+2\pi i\dot{\bf r}_{-1}[{\bf r}_{-1}\times{\bf r}_2]
-2\pi i\dot{\bf r}_1[{\bf r}_1\times{\bf r}_{-2}] +\dots
\end{equation}
If we neglect the internal degrees of freedom which describe deviations of 
the ring  from the ideal shape
$$
({\bf r}_{-1})^2=({\bf r}_1)^2=0 ,\qquad
{\bf r}_2={\bf r}_{-2}=0,\qquad\dots
$$
then the previous statement about canonically conjugated variables 
becomes obvious:
\begin{equation}
{\bf R}_n={\bf r}_{0n},\qquad
{\bf P}_n=2\pi \Gamma_n \cdot i[{\bf r}_{-1n}\times{\bf r}_{1n}]
\end{equation}
Such an approximation is valid only in the 
limit when  sizes of rings are small in comparison with the distances to 
the surface and the distances between different rings
\begin{equation}\label{dipoles}
\sqrt{\frac{P_n}{\Gamma_n}}\ll |Z_n|,|{\bf R}_{n}-{\bf R}_{l}|,
\qquad l\not=n. 
\end{equation}
These conditions are necessary for ensuring that the excitations of all internal degrees 
of freedom are small. Obviously, this is not true when a ring approaches 
the surface. In that case one should take into account also the internal 
degrees of freedom for the vortex lines. 

The inequalities (\ref{dipoles}) also imply that vortex rings in the limit under 
consideration are similar to  point magnet dipoles. 
This analogy is useful for calculation of the Hamiltonian for interacting 
rings. In the main approximation we may restrict the analysis by 
taking into account the dipole-dipole interaction only.

It should be mentioned that in some papers
(see e.g. \cite{Chorin} and references in that book) the discrete variables
identical to ${\bf R}_{n}$ and ${\bf P}_n$ are derived in a different way and 
referred as the vortex magnetization variables. 

In the expression for the Hamiltonian, several simplifications can be made.
Let us recall that for each moment of time it is possible to decompose 
the velocity field into two components
\begin{equation}
{\bf v}={\bf V}_0+\nabla\phi.
\end{equation}
Here the field ${\bf V}_0$ satisfies the following conditions
$$
(\nabla\cdot{\bf V}_0)=0,\qquad \mbox{curl}{\bf V}_0={\bf \Omega},\qquad
({\bf n}\cdot{\bf V}_0)|_{z=\eta}=0.
$$
The boundary value of the surface wave potential $\phi({\bf r})$ is 
$\psi({\bf r}_\bot)$. In accordance with these conditions 
the kinetic energy is decomposed into two parts and the Hamiltonian
of the fluid takes the form
\begin{equation}
{\cal H}=\frac{1}{2}\int_{z<\eta}{\bf V}_0^2d^3{\bf r}+
\frac{1}{2}\int\psi(\nabla\phi\cdot d{\bf S})+
\frac{g}{2}\int\eta^2d{\bf r}_{\bot}
\end{equation}
The last term in this expression is the potential energy 
of the fluid in the gravitational field.
If all vortex rings are far away from the surface then its deviation from 
the horizontal plane is small
\begin{equation}\label{smallangle}
|\nabla\eta|\ll 1,\qquad|\eta|\ll|Z_n|
\end{equation}
Therefore in the main approximation the energy of dipoles interaction with
the surface can be described with the help of so called "images". 
The images are vortex rings with parameters
\begin{equation}
\Gamma_n,\qquad
{\bf R}^*_n=(X_n,Y_n,-Z_n),\qquad {\bf P}^*_n=(P_{xn},P_{yn},-P_{zn})
\end{equation}

The kinetic energy for the system of point rings and their images is 
the sum of the self-energies of rings and the dipole-dipole interaction between 
them. The expression for the kinetic energy of small amplitude surface waves 
employs the operator $\hat k$ which multiplies Fourier-components of 
a function by the absolute value $k$ of a two-dimensional wave vector ${\bf k}$.
So the real Hamiltonian ${\cal H}$ is approximately equal to the simplified 
Hamiltonian $\tilde{\cal H}$
$$
{\cal H}\approx\tilde{\cal H}=
\frac{1}{2}\int(\psi\hat k \psi+g\eta^2)d{\bf r}_{\bot}
+\sum_n {\cal E}_n(P_n)+
$$
$$
+\frac{1}{8\pi}\sum_{l\not= n}
\frac{3(({\bf R}_n-{\bf R}_l)\cdot{\bf P}_n)
(({\bf R}_n-{\bf R}_l)\cdot{\bf P}_l)-
|{\bf R}_n-{\bf R}_l|^2({\bf P}_n\cdot{\bf P}_l)}
{|{\bf R}_n-{\bf R}_l|^5}+\quad
$$
\begin{equation}
+\frac{1}{8\pi}\sum_{l n}
\frac{3(({\bf R}_n-{\bf R}^*_l)\cdot{\bf P}_n)
(({\bf R}_n-{\bf R}^*_l)\cdot{\bf P}^*_l)-
|{\bf R}_n-{\bf R}^*_l|^2({\bf P}_n\cdot{\bf P}^*_l)}
{|{\bf R}_n-{\bf R}^*_l|^5}
\end{equation}

With the logarithmic accuracy the self-energy of a thin vortex ring is given 
by the expression
\begin{equation}
{\cal E}_n(P_n)\approx\frac{\Gamma_n^2}{2}
\sqrt{\frac{P_n}{\pi\Gamma_n}}
\ln\left(\frac{(P_n/\Gamma_n)^{3/4}}{A_n^{1/2}}\right)
\end{equation}
where the small constant $A_n$ is proportional to the conserved volume of 
the vortex tube forming the ring. This expression can easily be derived if we 
take into account that the main contribution to the energy is from the 
vicinity of the tube where the velocity field is approximately the same as near a 
straight vortex tube. The logarithmic integral  should then be taken between the 
limits from the thickness of the tube to the radius of the ring.

In the relation $\Psi=\Phi_0+\psi$ the potential $\Phi_0$ is approximately equal 
to the potential created on the flat surface by the dipoles and their images
\begin{equation}
\Phi_0({\bf r}_\bot)\approx\Phi({\bf r}_\bot)=
-\frac{1}{2\pi}\sum_n
\frac{({\bf P}_n\cdot({\bf r}_\bot-{\bf R}_n))}
{|{\bf r}_\bot-{\bf R}_n|^3}
\end{equation}

In this way we arrive at the following simplified system describing 
the interaction of point vortex rings with the free surface
\begin{equation}\label{tildeL}
\tilde{\cal L}=\sum_n \dot{\bf R}_n{\bf P}_n+
\int \dot\eta(\psi+\Phi)d^2{\bf r}_\bot-
\tilde{\cal H}[\{{\bf R}_n,{\bf P}_n\},\eta,\psi]
\end{equation}

It should be noted that due to the condition (\ref{dipoles}) the maximum
value of the velocity $V_0$ on the surface is much less then the typical 
velocities of the vortex rings
$$
\frac{P_n}{Z_n^3}\ll\frac{\Gamma_n^{3/2}}{P_n^{1/2}}
$$ 
Therefore the term $V_0^2/2$ in the Bernoulli equation 
$$
\Psi_t +V_0^2/2 +g\eta +\mbox{small corrections}=0
$$
is small in comparison with the term $\Psi_t$. 
The Lagrangian (\ref{tildeL}) is in accordance with this fact because it does 
not take into account terms like 
$(1/2)\int V_0^2\eta d^2{\bf r}_\bot$ in the Hamiltonian expansion.

\section{Interaction of the vortex ring with its image}

Now let us for simplicity consider the case of a single ring.
It is shown in the next section, 
that for a sufficiently deep ring the interaction with its image 
is much stronger than the interaction with the surface waves.
So it is interesting to examine the motion of the 
ring neglecting the surface deviation. 
In this case we have the integrable Hamiltonian for the system 
with two degrees of freedom
\begin{equation}\label{2df}
H=\frac{1}{64\pi}\left(
\alpha(P_x^2+P_z^2)^{1/4} -\frac{2P_z^2+P_x^2}{|Z|^3}\right),
\qquad Z<0
\end{equation}
where $\alpha\approx \mbox{const}$. The system has integrals of motion
$$
P_x=p=const, \qquad H=E=const
$$
so it is useful to consider the level lines of the energy function
in the left $(Z,P_z)$-half-plane taking $P_x$ as the parameter 
(see the Figure).
\begin{figure}
\label{E_levels}
\epsfxsize=360pt
\epsffile{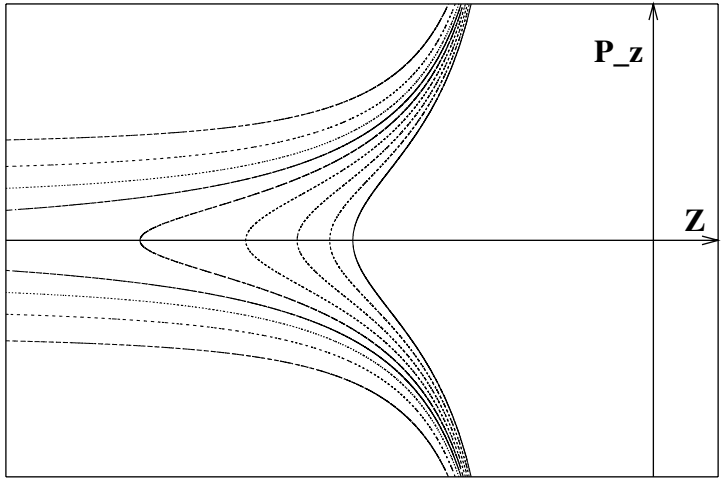}
\caption{ The sketch of level lines of the function 
$H(Z,P_z)$, Eq.(\ref{2df}).}
\end{figure}

One can distinguish three regions of qualitatively different behavior 
of the ring in that part of this half-plane where our approximation is valid
(see Eq.(\ref{dipoles})). 
In the upper region the phase trajectories come from infinitely
large negative $Z$ where they have a finite positive value of $P_z$. 
In the process of motion $P_z$ increases.
This behavior corresponds to the case when the ring approaches the surface. 
Due to the symmetry of the Hamiltonian (\ref{2df}) there is
a symmetric lower region, where the vortex ring moves away from the surface.
And there is the middle region, where $P_z$ changes the sign from negative to
positive at a finite value of $Z$. This is the region of the finite motion.

In all three cases the track of the vortex ring bends toward the surface, 
i.e. the ring is "attracted" by the surface.

\section{Cherenkov interaction of a vortex ring with surface waves}

When the ring is not very far from the surface and not very slow, 
the interaction with the surface waves becomes significant. 
Let us consider the effect of Cherenkov radiation of surface waves 
by a vortex ring which moves from the infinity to the surface. 
This case is the most definite from the viewpoint of initial conditions choice.
We suppose that the deviation of the free surface from the horizontal plane 
$z=0$ is zero at $t\to-\infty$, and we are interested in the asymptotic 
behavior of fields $\eta$ and $\psi$ at large negative $t$. 
In this situation we can neglect the interaction of the ring with its 
image in comparison with the self-energy 
and concentrate our attention on interaction with surface waves only.

The ring moves in the $(x,z)$-plane with an
almost constant velocity. In the main approximation the position 
${\bf R}$ of the vortex ring is given by the relations
\begin{equation}
{\bf R}\approx{\bf C}t,\qquad 
{\bf C}={\bf C}({\bf P})=
\frac{\partial{\cal E}}{\partial{\bf P}}=(C_x,0,C_z)\sim\frac{\bf P}{P^{3/2}},
\end{equation}
$$
\qquad C_x>0,\qquad C_z>0,\qquad t<0.
$$
The equations of motion for the Fourier-components of $\eta$ and $\psi$ follow
from the Lagrangian (\ref{tildeL})
\begin{equation}
\dot\eta_{\bf k}=k\psi_{\bf k},\qquad
\dot\psi_{\bf k}+g\eta_{\bf k}=-\dot\Phi_{\bf k}
\end{equation}
Eliminating $\eta_{\bf k}$ we obtain an equation for $\psi_{\bf k}$
\begin{equation}
\ddot\psi_{\bf k}+gk\psi_{\bf k}=-\ddot\Phi_{\bf k}
\end{equation}
where $\Phi_{\bf k}$ is the Fourier-transform of the  function 
$\Phi({\bf r}_\bot)$. Simple calculations give
$$
\Phi_{\bf k}=\frac{e^{-ik_x X}}{2\pi}\int
\frac{P_z Z-P_x x}{\sqrt{(x^2+y^2+Z^2)^3}}
e^{-i(k_x x+k_y y)}dx\,dy=
$$
\begin{equation}
=-\frac{e^{-ik_x X}}{2\pi}
\left(P_zD(k|Z|)+i\frac{P_x}{|Z|}\frac{\partial}{\partial k_x}D(k|Z|)
\right)
\end{equation}
where
\begin{equation}
D(q)=\int\frac{e^{-iq\alpha}d\alpha\,d\beta}
{\sqrt{(\alpha^2+\beta^2+1)^3}}=2\pi e^{-|q|}
\end{equation}
Finally, we have for $\Phi_{\bf k}$
\begin{equation}\label{Phi_k}
\Phi_{\bf k}=\left(\frac{iP_xk_x}{k}-P_z\right)e^{-k|Z|-ik_xX}=
\left(\frac{iP_xk_x}{k}-P_z\right)e^{t(kC_z-ik_xC_x)}
\end{equation}
Due to the exponential time behavior of $\Phi_{\bf k}(t)$ 
it is easy to obtain the expressions for 
$\psi_{\bf k}(t)$ and  $\eta_{\bf k}(t)$. Introducing the definition
\begin{equation}
\lambda_{\bf k}=kC_z-ik_xC_x
\end{equation}
we can represent the answer in the following form
\begin{equation}
\psi_{\bf k}(t)=\frac{-\left(\frac{iP_xk_x}{k}-P_z\right)\lambda_{\bf k}^2}
{gk+\lambda_{\bf k}^2}e^{\lambda_{\bf k}t}
=\left(\frac{P}{Ck}\right)\frac{\lambda_{\bf k}^3}
{gk+\lambda_{\bf k}^2}e^{\lambda_{\bf k}t}
\end{equation}
\begin{equation}\label{eta_kt}
\eta_{\bf k}(t)=\left(\frac{P}{C}\right)\frac{\lambda_{\bf k}^2}
{gk+\lambda_{\bf k}^2}e^{\lambda_{\bf k}t}
\end{equation}

The radiated surface waves influence the motion of the vortex ring. 
The terms produced by the field $\eta_{\bf k}(t)$ in the equations of motion 
for the ring come from the part $\int \dot\eta\Phi d^2{\bf r}_\bot$ 
in the Lagrangian (\ref{tildeL}). Using Eq.(\ref{Phi_k}) for 
the Fourier-transform of $\Phi$ we can represent these terms as follows
\begin{equation}
\delta \dot X=\int\frac{d^2{\bf k}}{(2\pi)^2}\dot\eta_{\bf k}
\frac{ik_x}{k}e^{kZ+ik_xX}
\end{equation}
\begin{equation}
\delta \dot Z=\int\frac{d^2{\bf k}}{(2\pi)^2}\dot\eta_{\bf k}
e^{kZ+ik_xX}
\end{equation}
\begin{equation}
\delta \dot P_x=-\int\frac{d^2{\bf k}}{(2\pi)^2}\dot\eta_{\bf k}\cdot(ik_x)
\left(P_z+\frac{iP_xk_x}{k}\right)e^{kZ+ik_xX}
\end{equation}
\begin{equation}
\delta \dot P_z=-\int\frac{d^2{\bf k}}{(2\pi)^2}\dot\eta_{\bf k}\cdot k
\left(P_z+\frac{iP_xk_x}{k}\right)e^{kZ+ik_xX}
\end{equation}
We can use Eq.(\ref{eta_kt}) to obtain  
the nonconservative corrections for time derivatives of the ring 
parameters from these expressions.
It is convenient to write down these corrections in the autonomic form
\begin{equation}\label{dX}
\delta \dot X=\left(\frac{P}{C}\right)\int\frac{d^2{\bf k}}{(2\pi)^2}
\left(\frac{ik_x}{k}\right)
\frac{(kC_z-ik_xC_x)^3}{gk+(kC_z-ik_xC_x)^2}
e^{-2k|Z|}
\end{equation}
\begin{equation}\label{dZ}
\delta \dot Z=\left(\frac{P}{C}\right)\int\frac{d^2{\bf k}}{(2\pi)^2}\cdot
\frac{(kC_z-ik_xC_x)^3}{gk+(kC_z-ik_xC_x)^2}
e^{-2k|Z|}
\end{equation}
\begin{equation}\label{dPx}
\delta \dot P_x=-\left(\frac{P}{C}\right)^2
\int\frac{d^2{\bf k}}{(2\pi)^2}\left(\frac{ik_x}{k}\right)
\frac{(kC_z-ik_xC_x)^2(C_z^2k^2+C_x^2k_x^2)}{gk+(kC_z-ik_xC_x)^2}e^{-2k|Z|}
\end{equation}
\begin{equation}\label{dPz}
\delta \dot P_z=-\left(\frac{P}{C}\right)^2
\int\frac{d^2{\bf k}}{(2\pi)^2}\cdot
\frac{(kC_z-ik_xC_x)^2(C_z^2k^2+C_x^2k_x^2)}{gk+(kC_z-ik_xC_x)^2}e^{-2k|Z|}
\end{equation}
where $C_x$ and $C_z$ can be understood as explicit functions of ${\bf P}$
defined by the dependence 
${\bf C}({\bf P})={\partial{\cal E}}/{\partial{\bf P}}$. 
More exact definition of $C_x$ and $C_z$ as $\dot X$ and $\dot Z$
is not necessary.

To analyze the above integrals let us first perform there the integration over 
the angle $\varphi$ in ${\bf k}$-space. 
It is convenient to use  the theory of contour integrals in the 
complex plane of variable $w=\cos\varphi$. The contour $\gamma$ of integration 
in our case goes clockwise just around the cut which is from $-1$ to $+1$. 
We define the sign of the square root $R(w)=\sqrt{1-w^2}$ so 
that its values are positive on 
the top side of the cut and negative on the bottom side. 
After introducing the quantities
\begin{equation}
a=\frac{C_z}{C_x},\qquad \omega_{\bf k}^2=gk,\qquad 
b_{\bf k}=\frac{\omega_{\bf k}}{C_x k}=\frac{1}{C_x}\sqrt{\frac{g}{k}}
\end{equation}
we have to use the following relations
$$
I_1(a,b)\equiv -\oint\limits_\gamma\frac{d\,w}{\sqrt{1-w^2}}\cdot
\frac{w(w+ia)^3}{b^2-(w+ia)^2}=\qquad\qquad\qquad\qquad\qquad\qquad\qquad
$$
\begin{equation}
=\pi(1+2b^2)+\pi i
\left(\frac{(b-ia)b^2}{\sqrt{1-(b-ia)^2}}-\frac{(b+ia)b^2}{\sqrt{1-(-b-ia)^2}}
\right)
\end{equation}

$$
I_2(a,b)\equiv i\oint\limits_\gamma\frac{d\,w}{\sqrt{1-w^2}}\cdot
\frac{(w+ia)^3}{b^2-(w+ia)^2}=\qquad\qquad\qquad\qquad\qquad\qquad\qquad
$$
\begin{equation}
=\pi\left(2a+\frac{b^2}{\sqrt{1-(b-ia)^2}}+\frac{b^2}{\sqrt{1-(-b-ia)^2}}
\right)
\end{equation}

$$
J_1(a,b)\equiv i\oint\limits_\gamma\frac{d\,w}{\sqrt{1-w^2}}\cdot
\frac{w(w+ia)^2(w^2+a^2)}{b^2-(w+ia)^2}=
\qquad\qquad\qquad\qquad\qquad\qquad
$$
\begin{equation}
=-4\pi ab^2+\pi\left(
\frac{b(b-ia)(a^2+(b-ia)^2)}{\sqrt{1-(b-ia)^2}}+
\frac{b(b+ia)(a^2+(b+ia)^2)}{\sqrt{1-(-b-ia)^2}}
\right)
\end{equation}

$$
J_2(a,b)\equiv \oint\limits_\gamma\frac{d\,w}{\sqrt{1-w^2}}\cdot
\frac{(w+ia)^2(w^2+a^2)}{b^2-(w+ia)^2}=
\qquad\qquad\qquad\qquad\qquad\qquad\quad
$$
\begin{equation}
=-2\pi(a^2+b^2+1/2)-\pi i\left(
\frac{b(a^2+(b-ia)^2)}{\sqrt{1-(b-ia)^2}}-
\frac{b(a^2+(b+ia)^2)}{\sqrt{1-(-b-ia)^2}}
\right)
\end{equation}
where the sign of the complex square root should be taken 
in accordance with the previous choice. 
It can easily be seen that the integrals $I_2$ and $J_1$ have resonance
structure at $a\ll 1$ and $|b|<1$. This is the Cherenkov effect itself.
Now the expressions (\ref{dX}-\ref{dPz}) take the form
\begin{equation}\label{dX2}
\delta \dot X=\frac{P_x}{(2\pi)^2}\int\limits_0^{+\infty}
I_1\left(a,b_{\bf k}\right)k^2e^{-2k|Z|}dk=
\frac{P_x}{(2\pi)^2}\left(\frac{g}{C_x^2}\right)^3 
F_1\left(a,\frac{2g|Z|}{C_x^2}\right)
\end{equation}
\begin{equation}\label{dZ2}
\delta \dot Z=\frac{P_x}{(2\pi)^2}\int\limits_0^{+\infty}
I_2\left(a,b_{\bf k}\right)k^2e^{-2k|Z|}dk=
\frac{P_x}{(2\pi)^2}\left(\frac{g}{C_x^2}\right)^3 
F_2\left(a,\frac{2g|Z|}{C_x^2}\right)
\end{equation}
\begin{equation}\label{dPx2}
\delta \dot P_x=\frac{P_x^2}{(2\pi)^2}\int\limits_0^{+\infty}
J_1\left(a,b_{\bf k}\right)k^3e^{-2k|Z|}dk=
\frac{P_x^2}{(2\pi)^2}\left(\frac{g}{C_x^2}\right)^4 
G_1\left(a,\frac{2g|Z|}{C_x^2}\right)
\end{equation}
\begin{equation}\label{dPz2}
\delta \dot P_z=\frac{P_x^2}{(2\pi)^2}\int\limits_0^{+\infty}
J_2\left(a,b_{\bf k}\right)k^3e^{-2k|Z|}dk=
\frac{P_x^2}{(2\pi)^2}\left(\frac{g}{C_x^2}\right)^4 
G_2\left(a,\frac{2g|Z|}{C_x^2}\right)
\end{equation}
Here the functions $F_1(a,Q)..G_2(a,Q)$ are defined by the integrals
\begin{equation}\label{F1}
F_1(a,Q)=\int\limits_0^{+\infty}I_1\left(a,\frac{1}{\sqrt{\xi}}\right)
\exp\left(-Q\xi\right)\xi^2\,d\xi
\end{equation}
\begin{equation}\label{F2}
F_2(a,Q)=\int\limits_0^{+\infty}I_2\left(a,\frac{1}{\sqrt{\xi}}\right)
\exp\left(-Q\xi\right)\xi^2\,d\xi
\end{equation}
\begin{equation}\label{G1}
G_1(a,Q)=\int\limits_0^{+\infty}J_1\left(a,\frac{1}{\sqrt{\xi}}\right)
\exp\left(-Q\xi\right)\xi^3\,d\xi
\end{equation}
\begin{equation}\label{G2}
G_2(a,Q)=\int\limits_0^{+\infty}J_2\left(a,\frac{1}{\sqrt{\xi}}\right)
\exp\left(-Q\xi\right)\xi^3\,d\xi
\end{equation}
and $Q={2g|Z|}/{C_x^2}$ is a dimensionless quantity
\footnote{
If we consider a fluid with surface tension $\sigma$, 
then two parameters appear: $Q$ and $T={g\sigma}/{C_x^4}$. 
In that case one should substitute $b_{\bf k}\to\sqrt{{1}/{\xi}+T\xi}$
as the second argument of the functions $I_1,I_2,J_1,J_2$ in the integrals
(\ref{F1}-\ref{G2})
}.
The Cherenkov effect is  most clear when
the motion of the ring is almost horizontal.
In this case $a\to +0$, and it is convenient to rewrite these
integrals without use of complex functions
\begin{equation}
F_1(+0,Q)=\pi\int\limits_0^{+\infty}\left(\xi^2+2\xi\right)
\exp\left(-Q\xi\right)\,d\xi-
2\pi\int\limits_0^{1}\frac{\xi\,d\xi}{\sqrt{1-\xi}}
\exp\left(-Q\xi\right)
\end{equation}
\begin{equation}
F_2(+0,Q)=G_1(+0,Q)=
-2\pi\int\limits_1^{+\infty}\frac{\xi^{3/2}\,d\xi}{\sqrt{\xi-1}}
\exp\left(-Q\xi\right)
\end{equation}
\begin{equation}
G_2(+0,Q)=-\pi\int\limits_0^{+\infty}\left(\xi^3+2\xi^2\right)
\exp\left(-Q\xi\right)\,d\xi+
2\pi\int\limits_0^{1}\frac{\xi^2\,d\xi}{\sqrt{1-\xi}}
\exp\left(-Q\xi\right)
\end{equation}
Here the square root is the usual positive defined real function. 
We see that only resonant wave-numbers contribute 
to the functions $F_2$ and $G_1$,
while  $F_1$ and $G_2$ are determined also by small values of $\xi$
which correspond to the large scale surface deviation  
co-moving with the ring.
So the effect of the Cherenkov radiation on the vortex ring motion
is the most distinct in the equations for $\dot Z$ and $\dot P_x$.
Especially it is important for $P_x$ because the radiation of surface waves
is the only reason for change of this quantity in the frame of our
approximation.

The typical values of $Q$ are large in practical situations. 
In this limit asymptotic values of the integrals above are
$$
F_1(+0,Q)\approx-\frac{9\pi}{2Q^4},\qquad 
G_2(+0,Q)\approx\frac{18\pi}{Q^5}
$$
$$
F_2(+0,Q)=G_1(+0,Q)\approx-2\pi\sqrt{\pi}\cdot\frac{\exp(-Q)}{\sqrt{Q}}
$$
and
$$
\delta \dot X\approx -\frac{9}{64\pi}\frac{P}{|Z|^3}\cdot\frac{1}{Q},\qquad 
\delta \dot Z\approx -\frac{1}{16\sqrt{\pi}}
\frac{P}{|Z|^3}\cdot Q^{2+1/2}\exp(-Q),\qquad 
$$
$$
\delta \dot P_x\approx  -\frac{1}{32\sqrt{\pi}}
\frac{P^2}{|Z|^4}\cdot Q^{3+1/2}\exp(-Q),\qquad 
\delta \dot P_z\approx +\frac{9}{32\pi}\frac{P^2}{|Z|^4}\cdot\frac{1}{Q}. 
$$
It follows from these expressions that the interaction with 
the surface waves is small in comparison with the 
interaction between ring and its image, if $Q\gg 1$. The corresponding 
small factors are $1/Q$ for $X$ and $P_z$, and $Q^{2+1/2}\exp(-Q)$ for $Z$.
As against the flat boundary, now $P_x$ is not conserved. 
It decreases exponentially slowly and this is the main effect of Cherenkov
radiation.

We see also that the interaction with waves turns the vector ${\bf P}$ 
towards the surface which results in a more fast boundary approach 
by the ring track.

\section{Conclusions and acknowledgments}

In this paper we have derived the simplified Lagrangian for the description
of the motion of deep vortex rings under free surface of perfect fluid. 
We have analyzed the integrable dynamics corresponding to the pure
interaction of the single point vortex ring with its image. It was found that
there are three types of qualitatively different behaviour of the ring. 
The interaction of the ring with the surface has an attractive
character in all three regimes.
The Fourier-components of radiated Cherenkov waves were calculated
for the case when the vortex ring comes from infinity and has both horizontal 
and vertical components of the velocity. The non-conservative corrections 
to the equations of motion of the ring, due to Cherenkov radiation, 
were derived. Due to these corrections the track of the ring bends towards
the surface faster then in the case of flat surface.
For simplicity, all calculations in Sec.5 were performed 
for a single ring. The generalization for the case of many 
rings is straightforward.

\medskip

The author thanks professor J.J. Rasmussen for his attention to this 
work and for helpful suggestions. 
This work was supported by the INTAS (grant No. 96-0413),
the Russian Foundation for Basic Research (grant No. 97-01-00093), 
and the Landau Postdoc Scholarship (KFA, Forschungszentrum, Juelich, Germany).

\end{document}